\documentclass[11pt,letterpaper]{article}
\usepackage[utf8]{inputenc}
\usepackage{multirow} 
\usepackage{booktabs} 
\usepackage[english]{babel}
\usepackage{amsmath}
\usepackage{amsfonts}
\usepackage{amssymb}
\usepackage{graphicx}
\usepackage[left=3cm,right=3cm,top=2cm,bottom=3cm]{geometry}
\usepackage[small,bf]{caption} 
\usepackage{framed} 
\usepackage{color} 
\usepackage{wrapfig}\definecolor{shadecolor}{RGB}{224,238,238} 

\author{Jorge Pinochet}
\title{\textbf{Rotating black holes: The most fantastic source of energy in the universe}}

\begin{document}

\renewcommand{\figurename}{\textbf{Figura}}

\author{Jorge Pinochet$^{*}$\\ \\
 \small{$^{*}$\textit{Facultad de Ciencias Básicas, Departamento de Física. }}\\
 \small{\textit{Universidad Metropolitana de Ciencias de la Educación,}}\\
 \small{\textit{Av. José Pedro Alessandri 774, Ñuñoa, Santiago, Chile.}}\\
 \small{e-mail: jorge.pinochet@umce.cl}\\}

\date{} 
\maketitle

\begin{center}\rule{0.9\textwidth}{0.1mm} \end{center}

\selectlanguage{english}

\begin{abstract}
\noindent Rotating black holes are the most powerful source of energy in the known universe, and are the cause of some of the most spectacular and extreme astronomical phenomena. The goal of this article is to analyze in simple terms the physics of energy extraction in rotating black holes. Specifically, the source of said energy, the efficiency of the energy extraction process, and some specific mechanisms that allow said extraction are analyzed. The article is intended primarily for undergraduate students of physics, astronomy and related fields. \\



\begin{center}\rule{0.9\textwidth}{0.1mm} \end{center}
\end{abstract}

\selectlanguage{english}

\maketitle

\section{Introduction}
According to Einstein's theory of general relativity, a black hole is a region of space-time whose gravity is so intense that nothing can escape from its interior, not even light. This definition seems to rule out any possibility of extracting energy from a black hole. However, as both theory and astronomical observation show, black holes are gravitational engines capable of converting their gravitational potential energy into enormous emissions of radiation, with an efficiency that can be dozens of times greater than the thermonuclear reactions that power stars like the Sun. These gravitational engines have two basic components: the central engine, which consists of a rotating black hole that drags along the space-time in its vicinity, and the fuel that powers the engine, which consists of the matter that swirls around it forming a disk-shaped structure composed of gas and dust called \textit{accretion disk}, which rotates at speeds close to the speed of light.\\

Energy extraction from rotating black holes is one of the most fascinating topics in modern astrophysics, but it is also a highly technical topic whose detailed description requires the use of the complex mathematics of general relativity [1,2], as well as other sophisticated physical tools and theories. However, as we will show in this paper, if a global view is sought, it is possible to understand the fundamentals of energy extraction in rotating black holes using only general physics and high school algebra. Therefore, this work can be useful as educational material in modern physics or undergraduate astronomy courses.\\

The paper is organized as follows. In Section 2, the concept of a Kerr black hole is introduced, which is a detailed mathematical model for a rotating black hole. In Section 3, the process of rotational energy extraction in a Kerr hole is examined in general terms, and a theoretical upper bound for this energy is established. In Section 4, two of the most popular mechanisms for extracting rotational energy in a black hole are discussed: the Penrose process and the Blandford-Znajek mechanism. The paper ends with a summary and some comments.

\section{Kerr black hole}
Black holes are the most extreme prediction of general relativity, which is the theory of gravity proposed by Einstein in 1915, which generalizes Newton's law of universal gravitation to the case of strong gravitational fields. In other words, Newtonian universal gravitation is a particular case of Einstein's theory that applies when gravity is weak. General relativity replaces the attractive gravitational forces of Newtonian physics with the concept of space-time curvature. Technically, a black hole is a type of exact solution in vacuum to the field equations of general relativity [3]. Black holes extremely curve space-time in their vicinity. Although physicists long refused to accept that black holes had a place in the universe, the evidence that has accumulated in their favor during the last decades is so numerous and compelling that no one seems to doubt that these objects are real anymore. \\

According to an important result of general relativity known as the \textit{no-hair theorem} [4], all solutions to the equations of Einstein's theory corresponding to a black hole can be characterized by only three externally observable classical parameters: mass $M$, electric charge $Q$, and angular momentum (rotation) $J$ [5]. The theorem assumes that spacetime is stationary and asymptotically flat. The latter means that the curvature gradually decreases as we move away from the black hole, so that, far from this object, gravity is very weak and space-time can be considered flat to a first approximation.

\begin{table}[htbp] 
\begin{center}
\caption{Main characteristics of the four types of black holes.}
\resizebox{0.5\textwidth}{!} {
\begin{tabular}{l l l} 
\toprule
\textbf{Property} & $J=0$ & $J \neq 0$ \\
\midrule
$Q = 0$ & Schwarzschild & Kerr \\

$Q\neq 0$ & Reissner-Nordstrom & Kerr-Newman \\

\midrule
\end{tabular}
}
\label{Main characteristics of the four types of black holes}
\end{center}
\end{table}

The word "hair" is used metaphorically to refer to observable characteristics of a black hole other than $M$, $J$, and $Q$, such as its chemical composition, magnetic field, and so on. Since information about the matter that makes up a black hole disappears after it is engulfed, it is inaccessible to an outside observer. As a result, all black holes with the same $M$, $J$, and $Q$ values look exactly the same—they are all "bald". The no-hair theorem implies that there are classically only four types of black holes, named after the physicists who found the corresponding mathematical solutions: (1) the Schwarzschild or static black hole, which depends only on $M$; (2) the Kerr black hole, which depends on $M$ and $J$; (3) the Reissner-Nordstrom black hole, which depends on $M$ and $Q$; and the most general, (4) the Kerr-Newman black hole, which depends on $M$, $J$, and $Q$. Table 1 summarizes the main characteristics of the four types of black holes allowed by the no-hair theorem. It is worth noting that these four types are stationary, meaning that their mathematical description is independent of time. \\

\begin{figure}[h]
\centering
\includegraphics[width=0.35\textwidth]{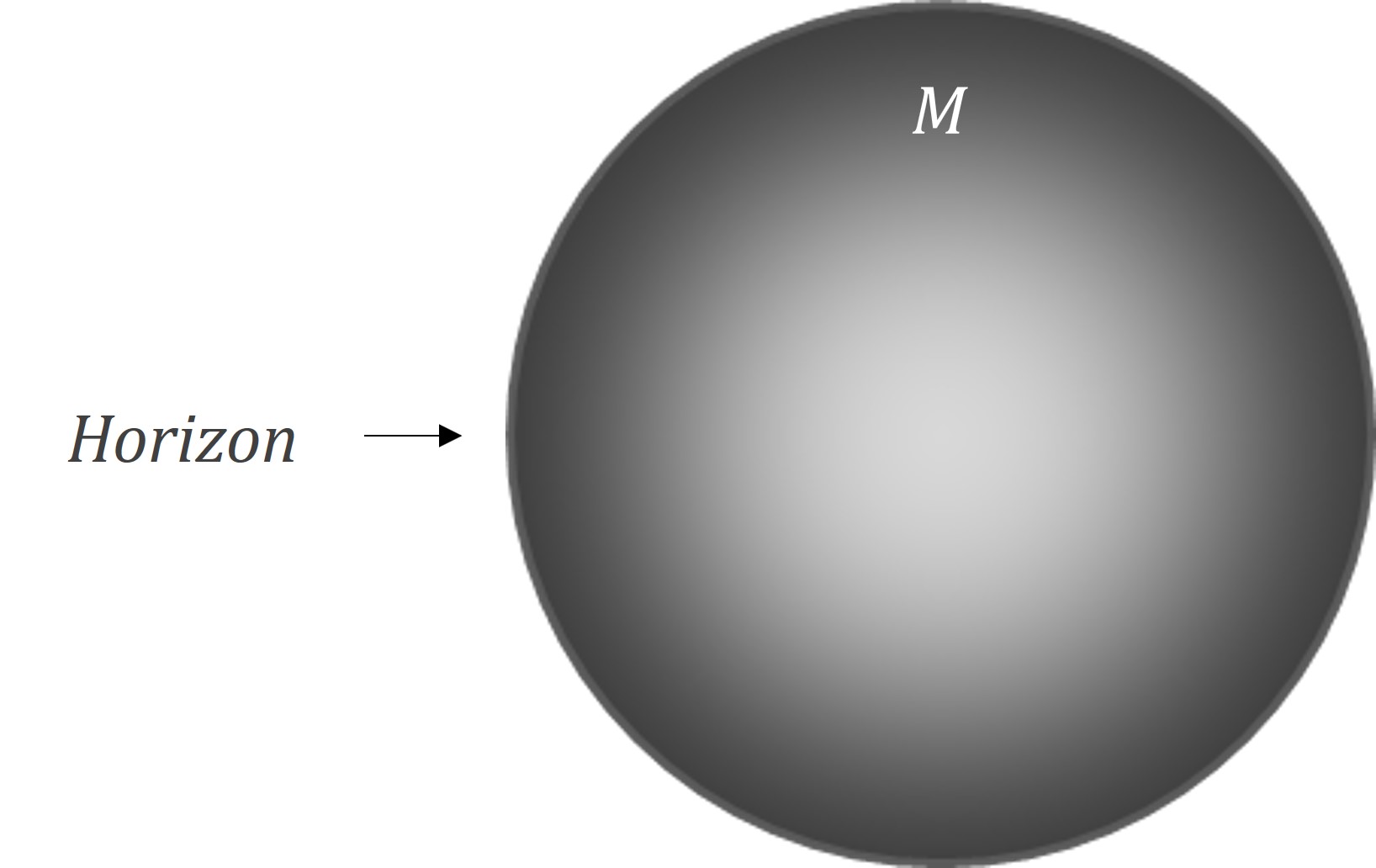}
\caption{Structure of a static black hole.}
\end{figure}

Although from a theoretical point of view the four types of black holes have equal status, from an astrophysical point of view the most important and realistic is the Kerr black hole. Indeed, the Reissner-Nordstrom and Kerr-Newman black holes are of more mathematical than astrophysical interest because electrical charges always tend to neutralize each other. Therefore, if a black hole has $Q \neq 0$, it will quickly attract opposite charges that will drive it to neutrality, thus becoming a Kerr or Schwarzschild black hole. However, because all objects in the universe rotate to some degree [6], and in particular, the stellar corpses from which the most common black holes in the universe form, also rotate, the Kerr solution provides the best mathematical model for describing the black holes observed by astronomers [5]. On the other hand, although the Schwarzschild black hole is also not very realistic due to its lack of rotation, if a Kerr black hole has a sufficiently small angular momentum, it can be described with a very good approximation by the static Schwarzschild solution. \\

\begin{figure}[h]
\centering
\includegraphics[width=0.6\textwidth]{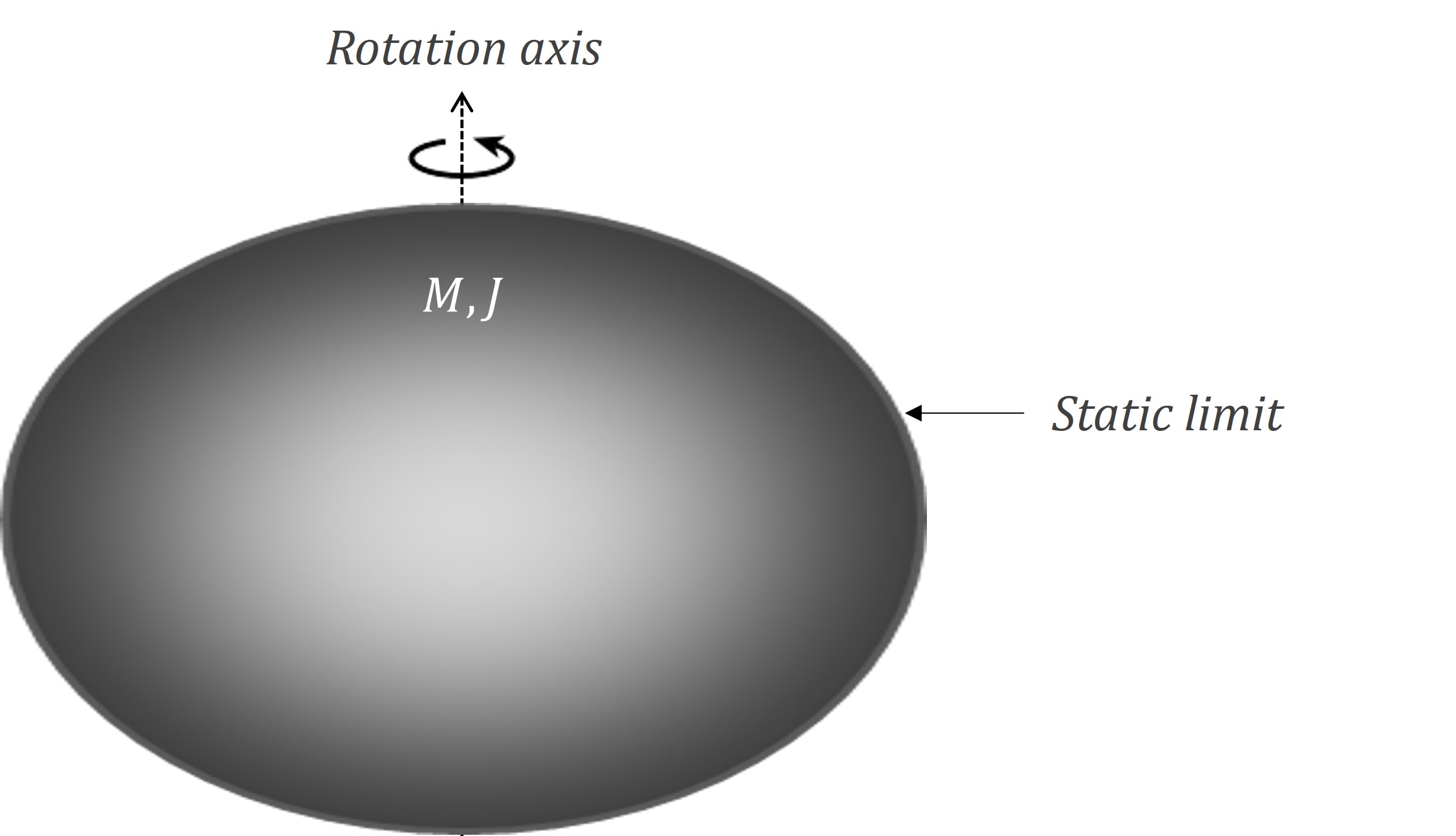}
\caption{Structure of a Kerr black hole seen parallel to its spin axis. The widening in the equatorial region is due to rotation.}
\end{figure}

For our purposes, the most important aspect of a Schwarzschild black hole is that when isolated, it represents a state of minimal energy. Using an analogy with the quantum properties of atoms and subatomic particles, a Schwarzschild black hole represents a ground state of minimal energy, whereas the other types of black holes, and in particular the Kerr black hole, can be thought of as excited states that emit energy as they decay to the ground state. Thus, once all the rotational energy has been extracted from a Kerr black hole, it stops spinning and becomes a Schwarzschild black hole, from which classically no more energy can be extracted. However, as Stephen Hawking showed in 1974–1975, when general relativity, quantum mechanics, and thermodynamics are combined, it is found that black holes emit thermal radiation and evaporate. These effects are only significant for hypothetical micro black holes. For astrophysical black holes, whose masses are greater than the mass of the Sun, the effects predicted by Hawking can be ignored, and the black hole can be described within the framework of classical general relativity. This impossibility of extracting energy from a Schwarzschild black hole was theoretically demonstrated by the Greek physicist Demetrios Christodoulou [7,8].\\

To analyze the generation of energy in rotating black holes, we need to describe the physical characteristics of these objects in more detail. As noted in the introduction, a black hole is a region of space-time whose gravity is so intense that nothing can escape from its interior, not even light. This region is delimited by a closed surface called \textit{horizon}, where the entire mass of the black hole is confined. Although the horizon is not a material surface, it can be imagined as a one-way membrane that only allows the flow of matter and energy inwards [9]. In the simple case of a static black hole, the horizon can be intuitively visualized as a spherical surface whose radius is given by\footnote{It is worth keeping in mind that, in general relativity, $R_{S}$ is a coordinate and not a physical distance to which a direct empirical meaning can be attributed.} [5,10]:

\begin{equation} 
R_{S}= \frac{2GM}{c^{2}},
\end{equation}

where $G=6.67 \times 10^{-11}N\cdot m^{2} \cdot kg^{-2}$ is the universal gravitational constant, $c= 3\times 10^{8}m \cdot s^{-1}$ is the speed of light in vacuum, $R_{S}$ is the \textit{Schwarzschild radius}, and $M$ is the mass of the black hole. According to general relativity, the mass $M$ is concentrated in a mathematical point of infinite density called \textit{singularity}. \\

The structure of a Kerr black hole is considerably more complex than that of a static black hole, since, first of all, it has two horizons, one inner and one outer. For a Kerr hole with angular momentum $J$, the outer horizon is located at [3,5]:

\begin{equation} 
R_{+} = \frac{GM}{c^{2}} + \sqrt{\frac{G^{2}M^{2}}{c^{4}} - \frac{J^{2}}{c^{2}M^{2}}}.
\end{equation}

The inner horizon is located in:

\begin{equation} 
R_{-} = \frac{GM}{c^{2}} - \sqrt{\frac{G^{2}M^{2}}{c^{4}} - \frac{J^{2}}{c^{2}M^{2}}}.
\end{equation}

It can be seen that $R_{+} \geq R_{-}$. These two equations can be summarized as:

\begin{equation} 
R_{\pm} = \frac{R_{S}}{2} \pm \sqrt{\frac{G^{2}M^{2}}{c^{4}} - \frac{J^{2}}{c^{2}M^{2}}}.
\end{equation}

where we have introduced Eq. (1). We see that for $J=0$, the Kerr hole becomes static, since $R_{-} = 0$, and $R_{+}$ becomes equal to the Schwarzschild radius that characterizes a static black hole ($R_{+} = R_{S}$). In the case of a Kerr black hole, the central singularity is not a mathematical point as it happens with the Schwarzschild black hole, but it is a ring located on the equatorial plane [5,10].\\

In addition to its two horizons, the Kerr black hole has a region outside the outer horizon called the \textit{ergosphere}, which has the property that no object inside it can remain static. Indeed, according to a phenomenon called the \textit{lens-thirring effect}, space-time in the vicinity of the Kerr black hole is pulled by its rotation, dragging all objects within the ergosphere along with it, preventing them from remaining static. A Kerr black hole is in perfectly rigid rotation; all points on the horizon have the same angular speed. [10,11] For this reason, the outer boundary of the ergosphere is known as the static limit (Fig. 2).\\ 

Because the ergosphere is outside the horizon, an object moving sufficiently rapidly outward can escape the black hole. The rotational energy stored in the ergosphere can also escape, which means it can be extracted, reaching colossal values in the case of the so-called \textit{extreme black hole}, which is one with the maximum possible rotation speed [5]. However, the extreme state is impossible to achieve in practice, although it is a useful mathematical model for calculating the energy that can be extracted from a Kerr black hole.\\

\begin{figure}[h]
\centering
\includegraphics[width=0.6\textwidth]{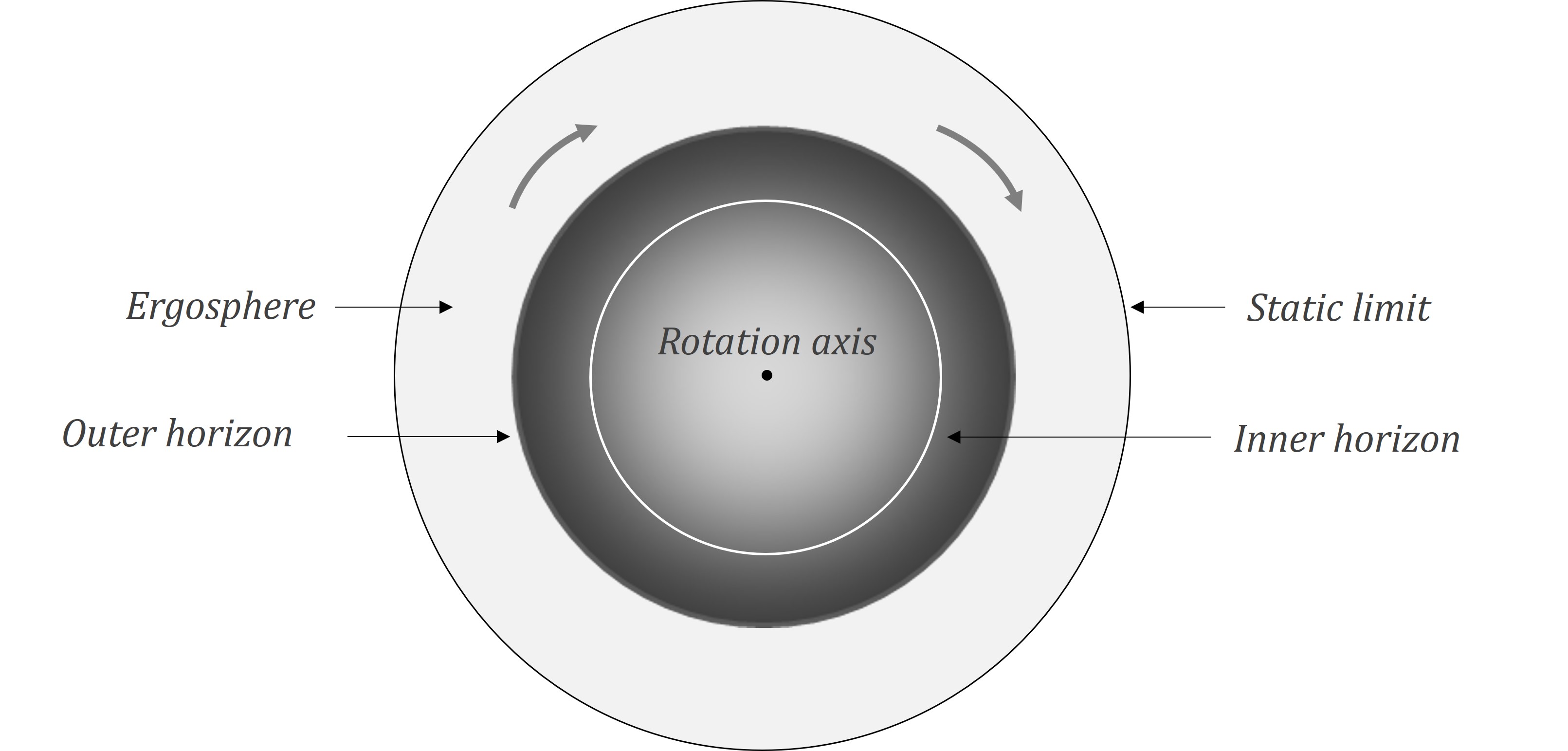}
\caption{A Kerr black hole seen from one of its poles.}
\end{figure}

What is the maximum angular momentum? The maximum value occurs when the square root in Eq. (4) is zero (note that this quantity cannot be less than zero, because it would become an imaginary number); imposing this condition results in:

\begin{equation} 
J_{max} = \frac{1}{2} McR_{S} = \frac{GM^{2}}{c}.
\end{equation}

When $J=J_{max}$ we say that the Kerr black hole is extreme. In energetic terms, an extreme hole is the opposite of a Schwarzschild black hole since, by definition, it corresponds to a state of minimum energy. Thus, the maximum energy that can be extracted occurs when an extreme Kerr black hole is reduced to a static or Schwarzschild one [3,5].\\

A key parameter to determine how much energy can be extracted from a Kerr hole is the area of the horizon. In the case of a static black hole, the relativistic calculation of the area coincides with the calculation made using elementary geometry:

\begin{equation} 
A_{S} = 4\pi R_{S}^{2} = \frac{16\pi G^{2} M^{2}}{c^{4}}.
\end{equation}

In the case of the Kerr black hole, the equation for the area is more complex, and elementary geometry no longer works. Calculations from general relativity show that the area of the outer horizon is given by [3,5]:

\begin{equation} 
A_{+} = \frac{8\pi G^{2} M^{2}}{c^{4}} \left( 1+ \sqrt{1- \left( \frac{cJ}{GM^{2}}\right)^{2} } \right) = \frac{8\pi G^{2} M^{2}}{c^{4}}\left( 1+ \sqrt{1- \left( \frac{J}{J_{max}}\right)^{2} } \right).
\end{equation}

where we have used Eq. (5). The area of the outer horizon is maximum when $J = 0$, which implies that the hole becomes static; under this condition, Eq. (6) agrees with Eq. (7). On the other hand, the area of the outer horizon is minimum when $J=J_{max}$, which implies that the black hole becomes extreme; imposing this condition in Eq. (7)\footnote{Comparing the equations for a static and a Kerr black hole, it is observed that $R_{S} > R_{+}$ and $A_{S} > A_{+}$.}:

\begin{equation} 
A_{+,min} = \frac{8\pi G^{2} M^{2}}{c^{4}}.
\end{equation}

In what follows, we will use the ideas developed in this section to analyze the rotational energy extraction in a Kerr black hole.

\section{Area theorem and energy extraction in Kerr holes}

The energy extracted from a Kerr black hole comes from its angular momentum $J$. As this energy is extracted, the rotational speed decreases. When it has completely exhausted its rotational energy, the Kerr hole becomes static. As we know, classically this is an irreducible final state, where no more energy can be extracted. Thus, the maximum rotational energy that can be extracted occurs when an extreme Kerr black hole becomes static (Fig. 4). In turn, this implies that the extreme black hole goes from having an area given by Eq. (8) to an area given by Eq. (6).\\

In the early 1970s, Stephen Hawking proved a theorem stating the following: \textit{The area of the horizon of a black hole can never decrease with time, so that after any physical process, the final area of the horizon must be greater than or at most equal to the initial area} [12] (the theorem can easily be extended to various black holes). Expressed mathematically, the theorem states that if $A_{i}$ is the initial area and $A_{f}$ is the final area, then it always holds that:

\begin{equation} 
A_{i} \leq A_{f}.
\end{equation}

This is the \textit{area theorem}, and by combining it with Eqs. (6) and (8), we can determine an upper bound on the energy that can be extracted from a Kerr black hole. Indeed, if we denote the initial mass of an extreme Kerr hole by $M_{i}$, and its final mass when all its rotational energy has been extracted and it has become static by $M_{f}$, then from Eqs. (6) and (8) we have:

\begin{equation} 
A_{i} = \frac{8\pi G^{2}M_{i}^{2}}{c^{4}}, \ A_{f} = \frac{16\pi G^{2}M_{f}^{2}}{c^{4}}.
\end{equation}

Applying Hawking's area theorem:

\begin{equation} 
\frac{8\pi G^{2}M_{i}^{2}}{c^{4}} \leq \frac{16\pi G^{2}M_{f}^{2}}{c^{4}}.
\end{equation}

Therefore:

\begin{equation} 
M_{i} \leq \sqrt{2}M_{f}.
\end{equation}

\begin{figure}[h]
 \centering
 \includegraphics[width=0.6\textwidth]{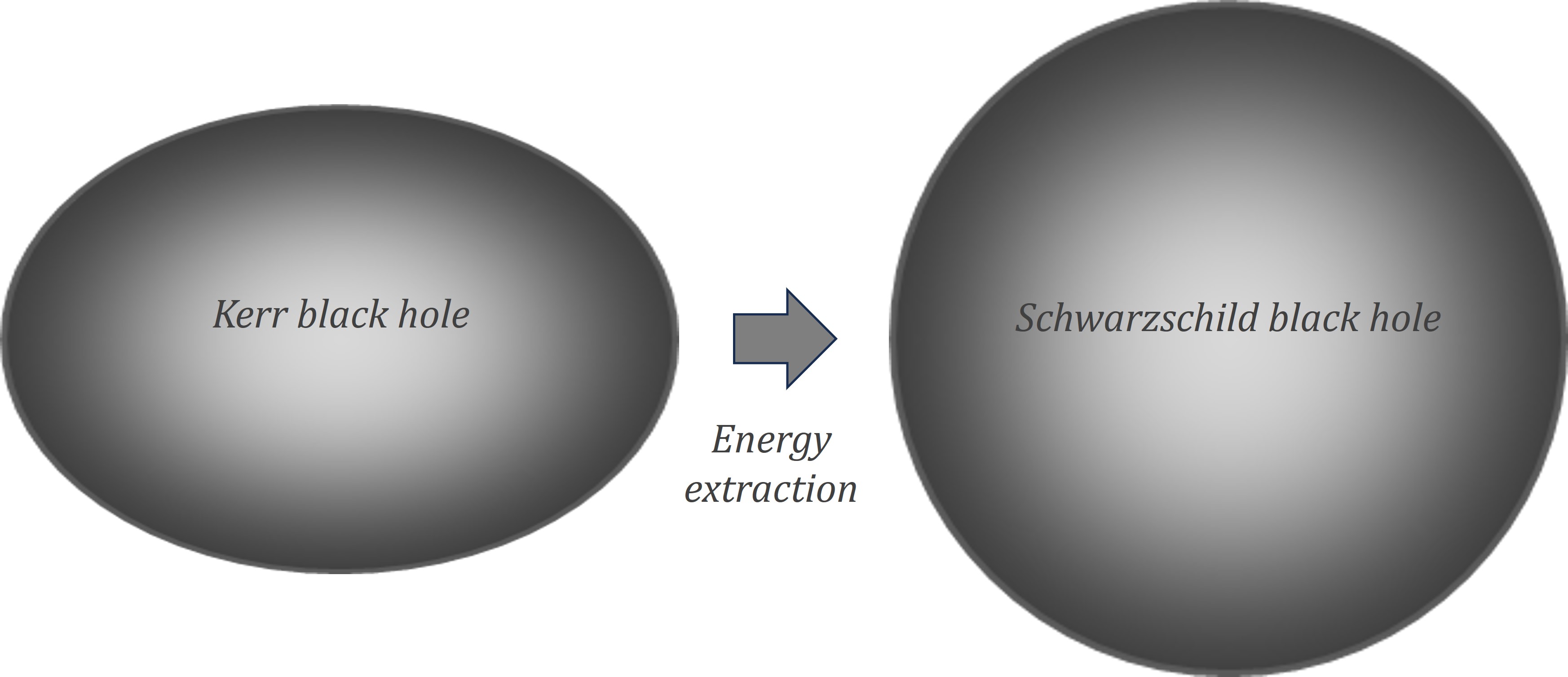}
 \caption{From the point of view of classical general relativity, a static black hole represents an irreducible final state.}
\end{figure}

To interpret this last result, remember that, according to the relativistic equivalence between mass and energy, rotational energy has an equivalent in mass, and by losing rotation, the hole must reduce its mass. From this idea, we can pose the following equation for the conservation of total energy:

\begin{equation} 
E_{rot} = M_{i}c^{2} - M_{f}c^{2} = \left( M_{i} - M_{f} \right)c^{2}.
\end{equation}

This equation tells us that the extracted rotational energy $E_{rot}$ is equal to the initial mass-energy, corresponding to an extreme hole, minus the final mass-energy, corresponding to a static hole. We can define the efficiency of the process as the quotient between the extracted energy and the initial energy:

\begin{equation} 
\varepsilon \equiv \frac{E_{rot}}{M_{i}c^{2}} = \frac{\left( M_{i} - M_{f} \right)c^{2}}{M_{i}c^{2}} = 1- \frac{M_{f}}{M_{i}}.
\end{equation}

Introducing Eq. (12):

\begin{equation} 
\varepsilon \equiv \frac{E_{rot}}{M_{i}c^{2}} \leq 1- \frac{M_{f}}{\sqrt{2}M_{f}} = 1- \frac{1}{\sqrt{2}} \cong 0.29
\end{equation}

This result tells us that the maximum efficiency of the process is $\varepsilon_{max} \cong 0.29$. In other words, the maximum rotational energy that can be extracted from an extreme Kerr black hole is 29\% of the initial mass-energy:

\begin{equation} 
E_{rot,max} = \varepsilon_{max} M_{i}c^{2} \cong 0.29 M_{i}c^{2} = 29\% M_{i}c^{2}.
\end{equation}

But this is an upper bound, so in practice, the extracted energy is lower. For a basis of comparison, remember that the efficiency of a nuclear fusion reaction is $\sim 1\%$. Note that Eqs. (15) and (16) represent a particular case, valid only for an extreme Kerr black hole. It can be shown that the general equation for the efficiency $\varepsilon$ as a function of angular momentum $J$ is [3,5]:

\begin{equation} 
\varepsilon = \frac{E_{rot}}{M_{i}c^{2}} = 1- \sqrt{\frac{1}{2}\left( 1+ \sqrt{1- \left( \frac{cJ}{GM^{2}}\right)^{2} } \right)}.
\end{equation}

From Eq. (5) we can rewrite this expression as:

\begin{equation} 
\varepsilon = 1- \sqrt{\frac{1}{2}\left( 1+ \sqrt{1- \left( \frac{J}{J_{max}}\right)^{2} } \right)} = 1- \sqrt{\frac{1}{2}\left( 1+ \sqrt{1- j^{2}} \right)}
\end{equation}

where we have defined the dimensionless quantity $j \equiv J/J_{max}$. For $j =1$, $J = J_{max}$ and $\varepsilon = \varepsilon_{max}$, so we recover Eq. (15). The graph of Eq. (18) appears in Figure 5. Due to the difficulties inherent in measuring the angular momentum $J$, estimates of $\varepsilon$ made by specialists are disparate, but most of them obtain an efficiency of a little less than 5\%, a figure that is far from the maximum limit given by Eq. (15), but which is much higher than the efficiency of nuclear reactions, and shows why Kerr black holes are associated with the most energetic phenomena in the universe.\\

\begin{figure}[h]
\centering
\includegraphics[width=0.6\textwidth]{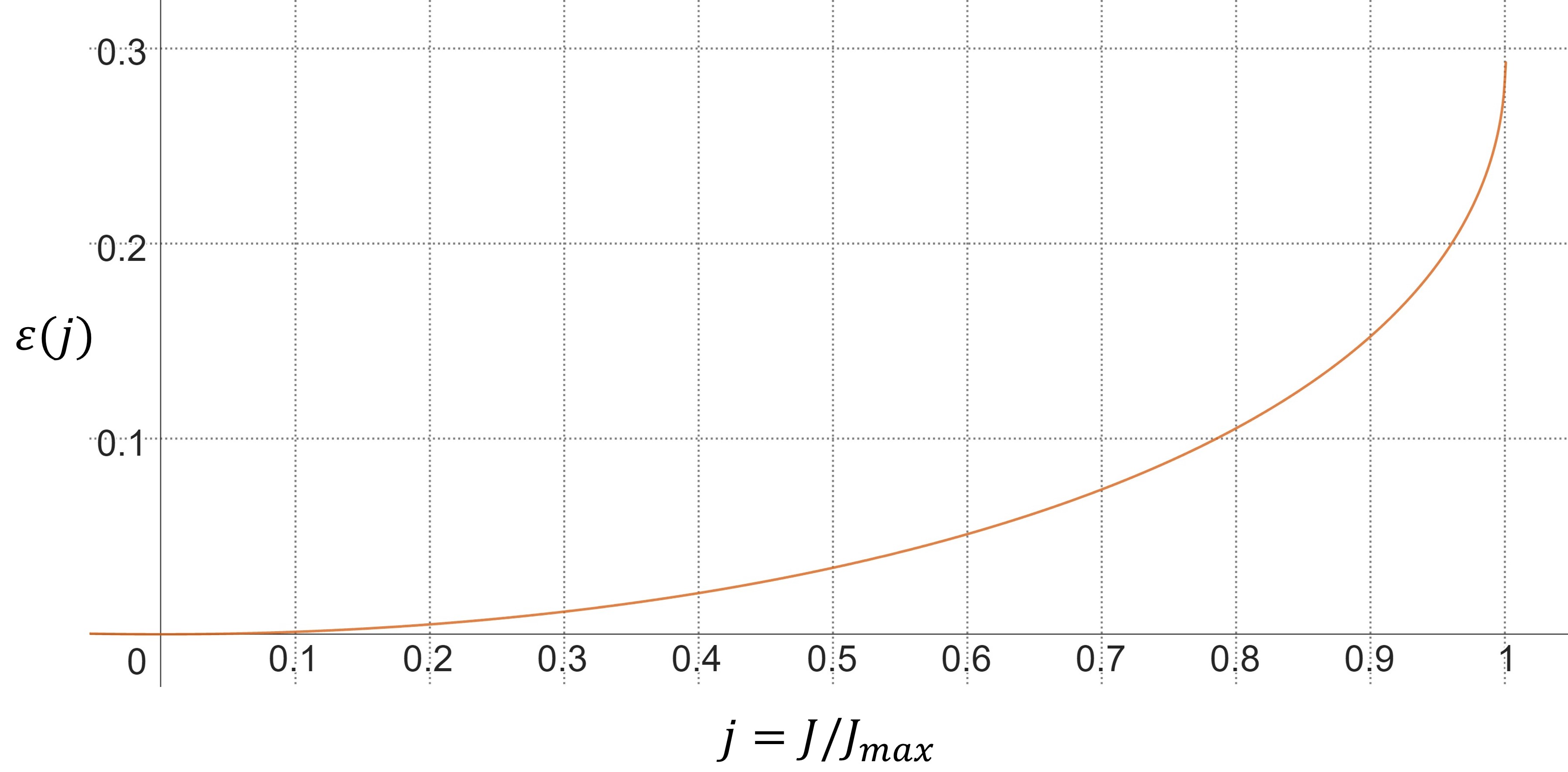}
\caption{Graph of efficiency ($\varepsilon$) vs dimensionless angular momentum ($j$) of a Kerr black hole. }
\end{figure}

As noted in the introduction, the most energetic phenomena in the universe, such as AGNs, are explained by a physical model where a rotating supermassive black hole, located at the center of a galaxy, acts as a gravitational engine fueled by an accretion disk. Since astronomical observations reveal that most supermassive black holes rotate at velocities close to $c$, which is the maximum value allowed by relativistic physics [13], the extreme Kerr black hole is a good mathematical model to explain the enormous energy emissions of AGNs.

\section{Energy extraction mechanisms}

So far we have discussed the process of energy extraction in a Kerr black hole in general terms, without considering any specific physical mechanism. Have such mechanisms been discovered? Do they have a place in the real universe? Several mechanisms have been proposed, and their astronomical applications are the subject of intense debate in current research. We will now discuss two of these mechanisms.

\subsection{The Penrose process}

One of the most popular mechanisms was proposed by Penrose in 1969 [14,15]. This mechanism allows a particle to be thrown into the ergosphere of a Kerr black hole and a fragment of this particle to return with greater energy than it initially had [16].\\
 
Before beginning the technical explanations, an analogy may be helpful. To understand the Penrose process, we can think of a carousel in an amusement park, which spins solely by inertia; that is, after receiving an initial impulse, it continues spinning even though the motors are off. If a child throws a ball at one of the carousel's horses, against the direction of rotation, the ball will bounce back, acquiring a greater speed than it initially had. As a result, the carousel will have lost some of its rotational speed. Thus, the ball's extra kinetic energy comes from the carousel's rotational kinetic energy. This analogy is useful, but, as we will see shortly, it is imperfect, among other reasons because the ball does not fragment.\\

To explain the Penrose process in more detail, let's imagine that we launch a particle from very far away into the ergosphere of a Kerr black hole, following a retrograde orbit, that is, a trajectory directed against the black hole's rotational direction. For simplicity, we can imagine that the retrograde orbit is contained within the black hole's equatorial plane. Suppose we calculate the trajectory so that upon entering the ergosphere, the particle fragments into two pieces, one of which is absorbed by the black hole, and the other escapes outward, moving an arbitrarily large distance away. Let $E_{inc}$ be the energy of the incident particle, $E_{esc}$ the energy of the escaping fragment, and $E_{abs}$ the energy of the absorbed fragment (Fig. 6). Due to the extreme intensity of gravity inside a black hole, general relativity allows the absorbed fragment to have negative energy.\footnote{The intuitive explanation for the negative energy of the absorbed particle is the following. The total energy of this particle is $E_{abs} = E_{p}+E_{c} + mc^{2}$, where $E_{p}$ is the (negative) potential energy, $E_{c}$ is the kinetic energy, and $mc^{2}$ is the energy associated with the rest mass of the particle of mass $m$. But, due to the extreme gravity of the black hole, inside the ergosphere, the potential energy is so negative that it exceeds the positive sum $E_{c} + mc^{2}$, that is, $E_{p}+E_{c}+mc^{2} <0$ so that the total energy of the absorbed particle is negative.}. Penrose showed that under these conditions it holds that:

\begin{equation}
E_{esc} > E_{inc}.
\end{equation}

That is, the escaping fragment has more energy than the incident particle.\footnote{It is interesting to note that in general relativity, the law of conservation of energy has no global physical meaning. One can only speak of the local conservation of energy, which is what happens in the Penrose process. Energy conservation can also be defined for asymptotically flat spacetime, as is the case with the Kerr solution, where the curvature is almost zero in regions very far from the black hole, making the Newtonian law of gravity a good approximation.}. We can demonstrate this curious effect through a very simple intuitive calculation. To do this, suppose that $Mc^{2}$ is the initial mass-energy of the black hole before absorbing the negative energy fragment. This means that the initial energy $E_{i}$ of the process is:

\begin{equation} 
E_{i} = Mc^{2} + E_{inc}.
\end{equation}

On the other hand, if $E_{abs}=-e$ is the energy of the absorbed fragment, the final energy $E_{f}$ of the process is:

\begin{equation} 
E_{f} = Mc^{2} + E_{abs} + E_{esc} = Mc^{2} -e + E_{esc}.
\end{equation}

where $Mc^{2} -e$ is the mass-energy of the black hole after absorption, which has been reduced by $e$. By conservation of energy, the initial energy of the process must equal the final energy, that is, $E_{i}=E_{f}$, so that, from Eqs. (20) and (21) we obtain $E_{inc}=-e+E_{esc}$ and therefore:

\begin{equation} 
E_{esc} = E_{inc} +e >E_{inc}.
\end{equation}

The trick to obtain this result is that the black hole absorbs negative energy, which leads to a reduction in its mass-energy, which translates into a decrease in its rotational speed. In other words, we have extracted rotational energy from the black hole.\\

\begin{figure}[h]
\centering
\includegraphics[width=0.6\textwidth]{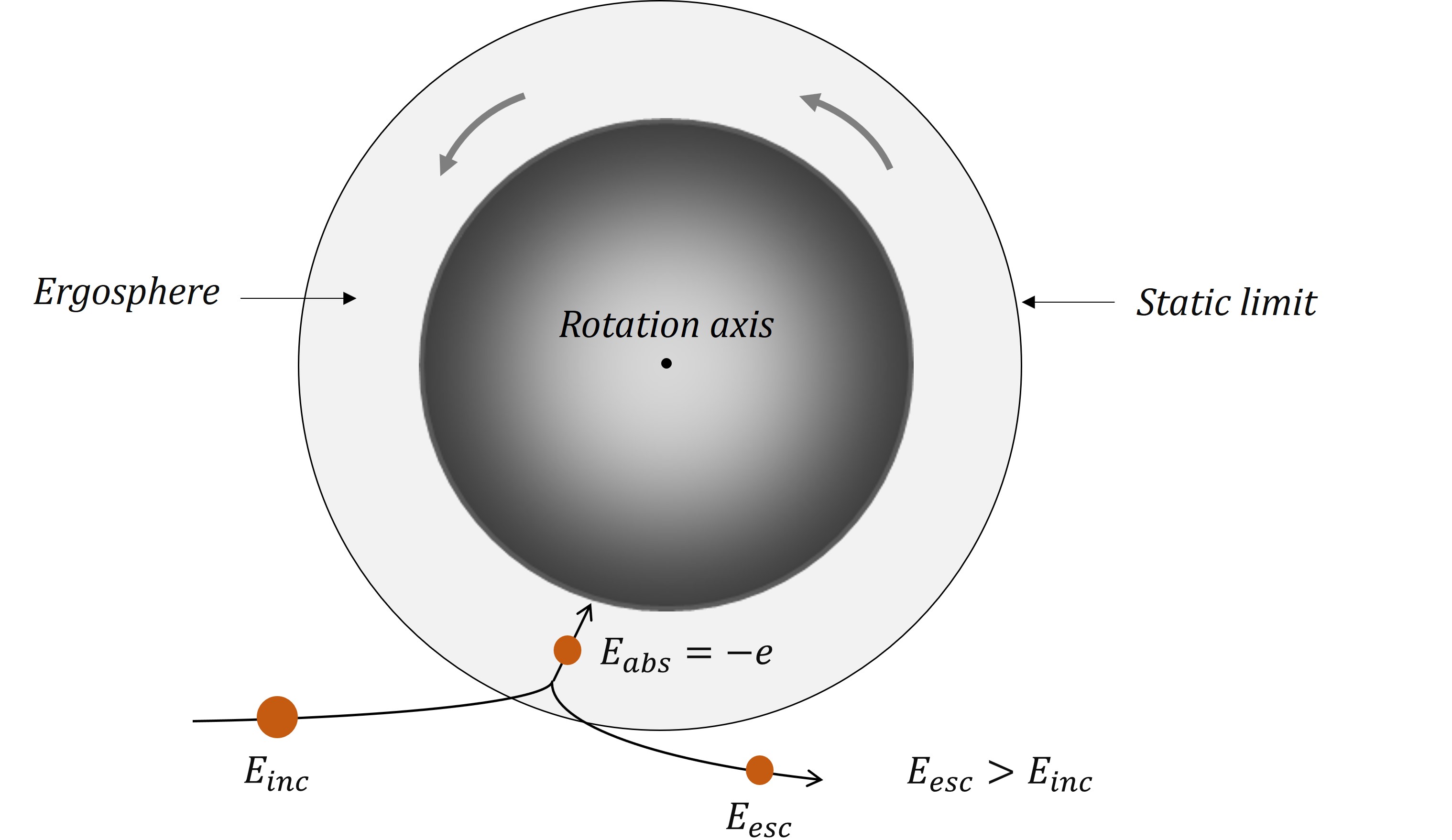}
\caption{The Penrose process seen from one of the poles of the Kerr black hole. Although the figure does not show the entire trajectory of the incident particle, it initially follows a retrograde orbit, but as it approaches the horizon, it is forced by the drag effect of the ergosphere to change the direction of its orbit, rotating in the same direction as the black hole.}
\end{figure}

According to Eq. (22), each time a particle with initial energy $E_{inc}$ is sent towards the ergosphere, the final energy extracted by the Penrose process is:

\begin{equation} 
\Delta E = E_{esc} - E_{inc} = e > 0.
\end{equation}

This energy is subject to the restrictions imposed by Eq. (15). This means that if the Penrose process is repeated a large number of times, and in each repetition an energy $\Delta E$ given by Eq. (23) is extracted, the maximum energy extracted cannot exceed 29\% of the initial mass-energy. Once this maximum figure is reached, the black hole will stop rotating and become static.\\

To define the efficiency of the Penrose process, Eq. (14) is impractical. It is more useful to define the efficiency as the ratio between the energy of the fragment escaping the ergosphere and the energy of the particle entering it:

\begin{equation} 
\varepsilon_{P} \equiv \frac{E_{esc}}{E_{inc}}.
\end{equation}

Since $E_{esc} > E_{inc}$, then $E_{esc}/E_{inc} > 1$, and $\varepsilon_{P}$ is always greater than 100\% . Detailed calculations by Robert Wald for an extreme Kerr hole reveal that the maximum efficiency for the Penrose process is $\varepsilon_{P,max} \approx 1.21$ [17]. In other words, the energy of the escaping fragment can be up to 21\% higher than the energy of the particle sent into the ergosphere. Although this efficiency may seem impressive, the Penrose process is actually impractical, requiring an enormous breakup velocity of the initial particle into its two fragments, as well as extremely high precision and synchronization. Furthermore, as Wald showed, it is impossible to obtain relativistic velocities (close to $c$) through the Penrose process [17], which makes it difficult to use it to explain highly energetic astrophysical phenomena, such as the AGN mentioned in the introduction.

\subsection{The Blandford-Znajek mechanism}

The limitations of the Penrose process have led some specialists to propose other mechanisms for extracting energy from the ergosphere. One of the most popular, known as the BZ mechanism, was suggested in 1977 by Roger Blandford and Roman Znajek [18]. As in the Penrose process, the extracted energy comes from the rotation of an extreme black hole; however, the BZ mechanism is more complex, as it assumes that the rotating black hole, in addition to being surrounded by an accretion disk, is immersed in a magnetic field oriented along its spin axis.\\

In general terms, the BZ mechanism is as follows. When the gas comprising the accretion disk falls toward the black hole, it describes a spiral trajectory. As it does so, its different parts rub against each other and heat up to temperatures of several million kelvins. These high temperatures ionize the gas, turning it into a plasma composed of a sea of positive ions and negative electrons. These churning charged particles generate turbulent magnetic fields, which channel relativistic plasma jets into two jets pointing in opposite directions, in the direction of the rotation axis (Fig. 7).
\\

\begin{figure}[h]
\centering
\includegraphics[width=0.4\textwidth]{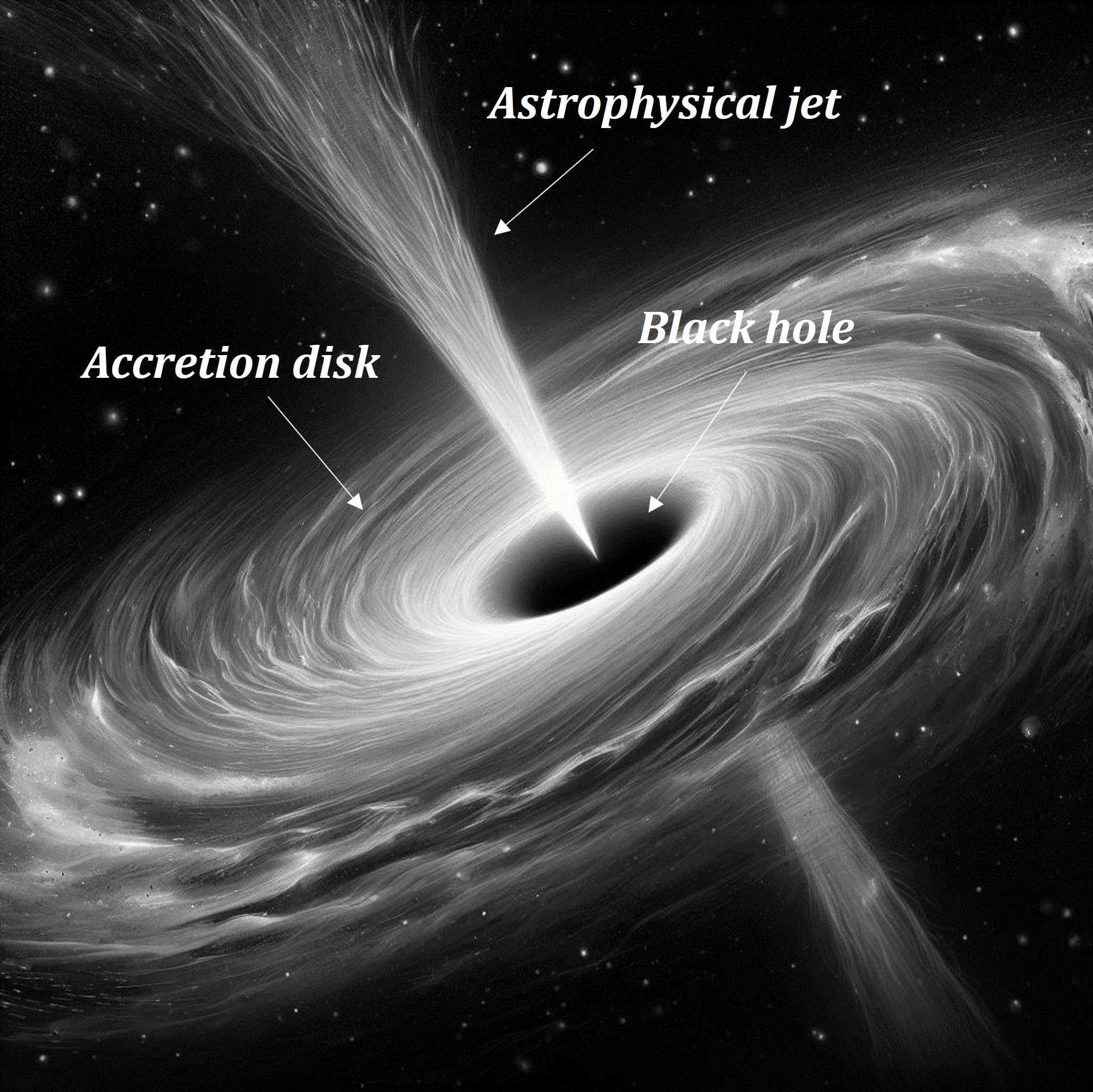}
\caption{Accretion disk and relativistic jets generated by a rotating black hole.}
\end{figure}

We can perform a heuristic calculation of the maximum power that can be extracted using the BZ mechanism. To do this, we need to analyze the mechanism in more detail. Imagine that a Kerr black hole is immersed in a magnetic field parallel to the rotation axis. Under these conditions, the field lines will pass through the hole. Since nothing can resist the rotation of space-time near the horizon, the field lines become anchored to the black hole and rotate with it, taking on a helix shape that opens up and down (Fig. 8).\\

Charged particles trapped in the magnetic field spin along with the induction lines, experiencing gigantic centrifugal forces. These forces cause the particles to move along the rotation axis, away from the black hole's poles, generating jets that can reach relativistic speeds, producing radiation that extends into the gamma ray domain. The energy transferred to the jets comes from the angular momentum of the Kerr black hole, and as this energy is emitted, the rotation speed is gradually reduced.\\

\begin{figure}[h]
\centering
\includegraphics[width=0.4\textwidth]{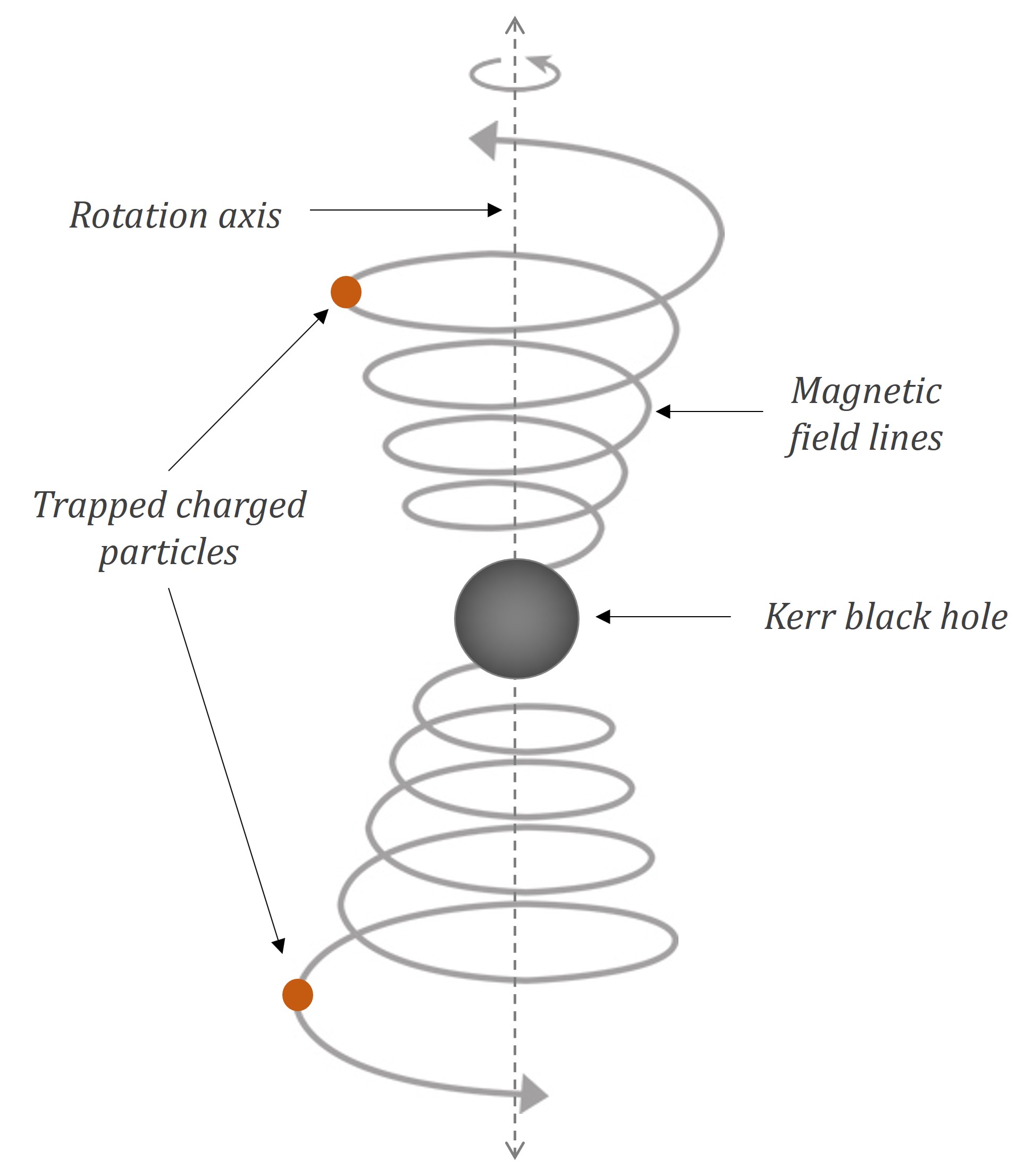}
\caption{A Kerr hole immersed in an external magnetic field.}
\end{figure}

To perform the heuristic calculation, let's start with a Kerr black hole of mass $M$. We can imagine this black hole as a solid sphere of radius $R_{+}$ with angular speed $\omega$ and moment of inertia $I$. Thus, it is possible to estimate the rotational kinetic energy of the black hole as:

\begin{equation} 
E_{R} \sim I\omega^{2} \sim MR_{+}^{2} \omega^{2}.
\end{equation}

Except for dimensionless factors, here we use the value of $I$ for a solid sphere of radius $R_{+}$. On the other hand, the total energy of the black hole, which we will assume includes the rotational energy, is:

\begin{equation} 
E_{T} \approx Mc^{2}.
\end{equation}

Next, we can calculate the fraction of the total energy that is rotational energy:

\begin{equation} 
\frac{E_{R}}{E_{T}} \sim \frac{R_{+}^{2}\omega^{2}}{c^{2}}.
\end{equation}

According to electromagnetic theory, the magnetic energy per unit volume is given by:

\begin{equation} 
E_{M,V} \sim \frac{B^{2}}{\mu_{0}}.
\end{equation}

where $B$ is the magnetic field strength, and $\mu_{0} = 4\pi \times 10^{-7} T \cdot m \cdot A^{-1}$ is the magnetic constant, also known as magnetic permeability. Since the black hole has radius $R_{+}$, the characteristic volume within which the magnetic energy is confined is $V \sim R_{+}^{3}$. Therefore, the black hole's magnetic energy can be estimated as:

\begin{equation} 
E_{M} \sim E_{M,V}V \sim \frac{B^{2}}{\mu_{0}} R_{+}^{3}.
\end{equation}

Thus, the energy extracted by the BZ mechanism is:

\begin{equation}
E_{BZ} \sim \left( \frac{E_{R}}{E_{T}}\right) E_{M} \sim \frac{B^{2} R_{+}^{5} \omega^{2}}{\mu_{0}c^{2}}. 
\end{equation}

Dimensionally, the characteristic time for the extraction of this energy is $t_{BZ} \sim R_{+}/c$, so we can estimate the net power as:

\begin{equation} 
E_{BZ} \sim \frac{E_{BZ}}{t_{BZ}} \sim \frac{B^{2} R_{+}^{4} \omega^{2}}{\mu_{0}c}.
\end{equation}

If the black hole is extreme, then the rotation speed of the outer horizon, which is of the order of $\omega R_{+}$, is close to the speed of light in vacuum, such that $\omega \sim c/R_{+}$. Plugging this expression into Eq. (31) gives the maximum power extracted:

\begin{equation} 
L_{BZ,max} \sim \frac{c}{\mu_{0}} B^{2}R_{+}^{2}.
\end{equation}

Detailed calculations reveal that the maximum power extracted by the BZ mechanism is:

\begin{equation} 
L_{BZ,max} \cong 4\pi \frac{c}{\mu_{0}} B^{2}R_{+}^{2}.
\end{equation}

Eq. (33) differs only by a dimensionless factor $4\pi$ from Eq. (32). If we assume that $R_{+}$ is of the order of the Schwarzschild radius, then $R_{+} \sim R_{S} \sim GM/c^{2}$. We can then rewrite Eq. (33) in terms of the black hole mass:

\begin{equation} 
L_{BZ,max} \cong \dfrac{4\pi G^{2}}{\mu_{0}c^{3}} B^{2}M^{2}.
\end{equation}

As an example, consider a supermassive black hole of $10^{8} M_{\odot}$, where $M_{\odot} \sim 10^{30} kg$ is the solar mass. If $B = 1T$, this results in $L_{BZ,max} \sim 10^{37} W$. In perspective, this energy, emitted in a single second, is greater than that consumed by the entire Earth in a year.\\

The BZ mechanism is one of the most accepted to explain the energy emission of AGNs. However, some specialists have suggested that a detailed explanation requires combining the Penrose process and the BZ mechanism [19]. In any case, it is important to remember that neither the Penrose process, nor the BZ mechanism, nor any other method of energy extraction can exceed the 29\% efficiency imposed by Eq. (15). Once this limit is reached, the black hole stops rotating and becomes static.

\section{Final comments}

The extraordinary energy efficiency that Kerr holes can achieve has led some authors to suggest a speculative but fascinating idea: the possibility of a highly advanced civilization using a rotating black hole as an energy source. It was Penrose himself in his 1969 paper who first proposed this idea [14]. Subsequently, other authors have suggested similar ideas [20]. This leads us to think that perhaps, in the distant future, our civilization could find in Kerr black holes a clean and efficient solution to the complex energy problems that we will surely have to face as a society, provided that we can survive our current technological infancy.\\

Although Kerr black holes are objects of scientific interest in their own right, one of their most attractive aspects is their energy efficiency, which allows them to explain the most extreme phenomena in the universe. For this reason, the extraction of energy in Kerr black holes is a hot topic of current research, both from a theoretical and observational perspective. In particular, the Penrose process and the BZ mechanism continue to generate questions that are actively investigated. This reveals the breadth and complexity of the topic we have addressed and the impossibility of covering it in detail in an article such as this, intended for non-specialists. \\

In any case, it is hoped that the reader has found in this work a stimulating, accessible and up-to-date exposition, which encourages him to continue to delve deeper into this and other topics related to the physics of black holes.

\section*{References}

[1]	M. Camenzind, Compact Objects in Astrophysics: White Dwarfs, Neutron Stars and Black Holes, Springer, Berlin, 2007.

\vspace{2mm}

[2]	D. Maoz, Astrophysics in a Nutshell, Princeton University Press, Princeton, 2007.

\vspace{2mm}

[3]	V.P. Frolov, A. Zelnikov, Introduction to Black Hole Physics, Oxford University Press, Oxford, 2011.

\vspace{2mm}

[4]	R. Ruffini, J.A. Wheeler, Introducing the black hole, Physics Today 24 (1971) 30–41.

\vspace{2mm}

[5]	V.P. Frolov, I.D. Novikov, Black Hole Physics: Basic Concepts and New Developments, Springer Science, Denver, 1998.

\vspace{2mm}

[6]	K.L. Lang, Essential Astrophysics, Springer, Berlin, 2013.

\vspace{2mm}

[7]	D. Christodoulou, Reversible and Irreversible Transformations in Black-Hole physics, Physical Review Letters 25 (1970) 1596–1597.

\vspace{2mm}

[8]	D. Christodoulou, Reversible Transformations of a Charged Black Hole, Physical Review D 4 (1971) 3552–3555.

\vspace{2mm}

[9]	S.W. Hawking, Historia del tiempo: Del big bang a los agujeros negros, Crítica, Barcelona, 1988.

\vspace{2mm}

[10] J.P. Luminet, Agujeros negros, Alianza, Madrid, 1991.

\vspace{2mm}

[11] B. Schutz, Gravity from the Ground Up, Cambridge University Press, Cambridge, 2003.

\vspace{2mm}

[12] S.W. Hawking, Black Holes in General Relativity, Communications in Mathematical Physics 25 (1972) 152–166.

\vspace{2mm}

[13] C.S. Reynolds, Measuring Black Hole Spin Using X-Ray Reﬂection Spectroscopy, Space Science Reviews 183 (2014) 277–294.

\vspace{2mm}

[14] R. Penrose, Gravitational collapse: the role of general relativity, Nuovo Cimento 1 (1969) 252–276.

\vspace{2mm}

[15] R. Penrose, R.M. Floyd, Extraction of Rotational Energy from a Black Hole, Nature 229 (1971) 177–179.

\vspace{2mm}

[16] R.M. Wald, Space, Time, and Gravity: The Theory of the Big Bang and Black Holes, The University of Chicago Press, Chicago, 1992.

\vspace{2mm}

[17] R.M. Wald, Energy Limits on the Penrose Process, The Astrophysical Journal 191 (1974) 231–234. https://doi.org/10.1086/152959.

\vspace{2mm}

[18] R.D. Blandford, R.L. Znajek, Electromagnetic extraction of energy from Kerr black holes, Monthly Notices of the Royal Astronomical Society 179 (1977) 433–456. https://doi.org/10.1093/mnras/179.3.433.

\vspace{2mm}

[19] K. Parfrey, A. Philippov, B. Cerutti, First-Principles Plasma Simulations of Black-Hole Jet Launching, Phys. Rev. Lett. 122 (2019) 035101. https://doi.org/10.1103/PhysRevLett.122.035101.

\vspace{2mm}

[20] K.S. Thorne, Agujeros negros y tiempo curvo: El escandaloso legado de Einstein, Crítica, Barcelona, 2000.

\end{document}